\documentclass{optica-article}

\journal{opticajournal} 

\articletype{Research Article}

\begin{document}

\title{Ptychographic lensless coherent endomicroscopy through a flexible fiber bundle} 

\author{
Gil Weinberg, \authormark{1} 
Munkyu Kang, \authormark{2,3}
Wonjun Choi, \authormark{2,3}
Wonshik Choi, \authormark{2,3} 
and Ori Katz \authormark{1,*}
}

\address{

\authormark{1}Institute of Applied Physics, The Hebrew University of Jerusalem, 9190401 Jerusalem, Israel\\
\authormark{2}Center for Molecular Spectroscopy and Dynamics, Institute for Basic Science (IBS), Seoul 02481, Republic of Korea\\
\authormark{3}Department of Physics, Korea University, Seoul 02481, Republic of Korea}

\email{\authormark{*} Orik@mail.huji.ac.il} 

\begin{abstract*} %~100 words
    Conventional fiber-bundle-based endoscopes allow minimally invasive imaging through flexible multi-core fiber (MCF) bundles by placing a miniature lens at the distal tip and using each core as an imaging pixel. In recent years, lensless imaging through MCFs was made possible by correcting the core-to-core phase distortions pre-measured in a calibration procedure. However, temporally varying wavefront distortions, for instance, due to dynamic fiber bending, pose a challenge for such approaches. 
    Here, we demonstrate a coherent lensless imaging technique based on intensity-only measurements insensitive to core-to-core phase distortions.
    We leverage a ptychographic reconstruction algorithm to retrieve the phase and amplitude profiles of reflective objects placed at a distance from the fiber tip, using as input a set of diffracted intensity patterns reflected from the object when the illumination is scanned over the MCF cores. Our approach thus utilizes an acquisition process equivalent to confocal microendoscopy, only replacing the single detector with a camera. 
   \end{abstract*}

\section{Introduction}
	
Flexible fiber micro-endoscopes allow microscopic imaging deep inside tissues beyond conventional microscopes' absorption and scattering limits. Recent devices based on flexible optical fibers allow imaging with micrometer resolution, making them suitable for imaging neurons and cells in various contrast mechanisms \cite{flusberg2005fiber, oh2013optical,accanto2022flexible}. To generate an image, fiber-based endoscopes traditionally required bulky optical elements, such as lenses \cite{knittel2001endoscope} and/or mechanical scanners \cite{giniunas1993endoscope} to be placed at the distal fiber end, increasing their footprint and the potential tissue damage. 

While lensless endoscopy is a desired goal, the main hurdle in lensless coherent imaging through flexible fibers is the dynamic wavefront distortions due to phase randomization between the different transverse modes. 
In recent years, lensless fiber endoscopes utilizing bare multi-mode fibers (MMF) or multi-core fibers (MCF, also referred to as coherent fiber-bundles) have been demonstrated via wavefront-shaping techniques  \cite{bertolotti2022imaging, oh2023review}.  
In these works, the inherent wavefront distortions are corrected physically (using a spatial light modulator, SLM) or computationally. Initially, the wavefront distortions have been measured prior to the imaging experiments \cite{vcivzmar2012exploiting, choi2012scanner,ploschner2015seeing, caravaca2017single, li2021memory, thompson2011adaptive, tsvirkun2016widefield, gordon2019full, scharf2020video, wen2023single}, and the correction was effective as long as the fibers remained static after the wavefront distortion measurement (or calibration) step. 

However, as these calibration-based approaches cannot handle the temporally varying wavefront distortions of flexible fiber bending, more recent developments have allowed calibration-free lensless endoscopy based on either iterative wavefront optimization of a nonlinear signal or image metric\cite{weiss2018two, yeminy2021guidestar}, the addition of a distal mirror\cite{badt2022real}, computational reconstruction using multiple holographically measured fields \cite{choi2022flexible, kang2023fourier, haim2023image}, or phase-retrieval of simple incoherent targets \cite{porat2016widefield, stasio2016calibration}. 

Here, we demonstrate that coherent phase-sensitive lensless imaging through MCFs can be achieved by simple intensity measurements, without correcting the core-to-core phase distortions, by employing ptychographic phase retrieval.
Specifically, we implement a conventional raster-scanned illumination in epidetection geometry as performed in confocal-endoscopy \cite{gmitro1993confocal,knittel2001endoscope} to record the diffraction intensity patterns from targets placed at a small distance from the MCF (Fig.~\ref{fig1}).
This set of diffraction patterns, measured at the proximal facet of the fiber, is used as the input of a standard ptychographic reconstruction algorithm \cite{maiden2017further, pham2019semi}, which recovers the object amplitude and phase profile. 
The approach is agnostic to the MCF phase distortions as the MCF is used in detection only as an intensity relay, and by providing the illumination to the object through a single (scanned) core. The true phases of the diffraction patterns are computationally retrieved in the ptychographic reconstruction process.

\section{Methods}

\subsection{Principle}
    
%%%%%%%%%%%%%%% Figure 1 - Setup %%%%%%%%%%%%%%%%%%%%%%
\begin{figure}[ht!]
	\centering
	\includegraphics [width=\textwidth,]
	{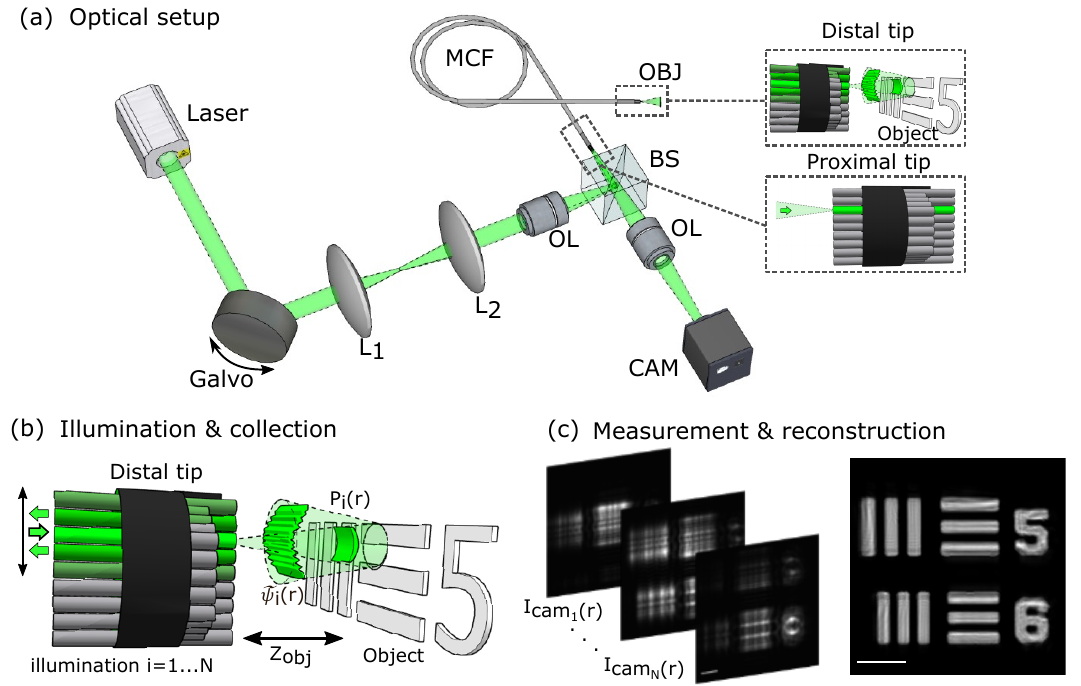}
	\caption{
		\textbf{Ptychographic lensless fiber-bundle endomicroscopy.} (a) The optical setup: A 2-axis galvanometer scans the illumination across the proximal facet of a multi-core fiber (MCF), illuminating each core one by one. An sCMOS camera that images the proximal facet records the diffraction patterns back-scattered from a target object placed at a small distance from the MCF distal tip for each illuminated core (inset). (b) At the $i$-th core illumination, a different area in the target object is illuminated, effectively realizing the scanning 'probe' of conventional ptychography, $P_i(\textbf{r})$. (c) Sample numerically simulated diffraction intensity patterns recorded for different illuminated cores (left) and the resulting reconstructed target intensity profile (right). Scale bars: 25$\mu m$.
        	}
	\label{fig1}
\end{figure}

The principle of ptychographic endoscopy is depicted in Fig.~\ref{fig1}. Our approach is based on the insight that for a target placed at a small distance from the fiber, conventional raster-scanned illumination of the fiber cores (Fig.~\ref{fig1}a inset), results in scanning a small illumination spot over the target area (Fig.~\ref{fig1}b). The reflected light intensity pattern measured at the proximal MCF facet for each illumination position is the diffraction pattern of each illuminated area on the target, which overlaps with the neighboring illuminated areas (Fig.~\ref{fig1}b). These measurements are thus equivalent to the measured diffraction patterns of ptychographic coherent-diffractive imaging (CDI) \cite{miao2000possible},  where a small illumination 'probe' is sequentially scanned over the target. The well-established ptychographic phase-retrieval algorithms \cite{rodenburg2019ptychography} can thus be utilized for reconstructing the complex-valued target from this set of lensless, intensity-only measurements (Fig.~\ref{fig1}c).

The process of ptychographic endoscopy thus consists of two steps:
1) Measurement step (Fig.~\ref{fig1}a,b): In every acquisition step, the illumination is focused on a single core, as shown in the inset of Fig.~\ref{fig1}a. This effectively illuminates a small area on the object, defined by the diffraction of the excited mode in the single core. The reflected light from the illuminated object area propagates back to the distal MCF facet (Fig.~\ref{fig1}b), and the resulting diffraction pattern is collected by the MCF and recorded in its proximal end. Scanning the illumination over the MCF cores results in overlapping illuminated areas across the object.
2) Reconstruction step (Fig.~\ref{fig1}c): The camera images of the diffraction patterns from each of the illuminated cores (Fig.~\ref{fig1}c, left) are fed into a ptychographic reconstruction algorithm.
In addition to the recorded intensity patterns, an initial estimation of the illumination (the ptychographic 'probe') and the distance from the fiber to the target are all fed as inputs to the reconstruction algorithm. The standard ptychographic iterative reconstruction algorithm searches for the complex-valued object (and exact probe function) that will best fit the entire set of measured diffraction intensity patterns, i.e., it retrieves the amplitude and phase of the target object (Fig.~\ref{fig1}c, right) \cite{rodenburg2019ptychography}.

\subsection{Forward model and reconstruction algorithm}
Here, we present a straightforward mathematical model of the forward process and the reconstruction algorithm. Consider a monochromatic coherent illumination given by the complex-valued electric field distribution at the MCF distal facet, $E_i(\textbf{r})$, where $\textbf{r}$ is the transverse position along the MCF facet, and $i$ is an index marking the illuminated core. The illumination function at the object plane (the 'probe' function)  is given by the free-space propagation of $E_i(\textbf{r})$ to the target plane: $P_{i}(\textbf{r}) = \mathcal{P}_{z}{[E_i(\textbf{r})]}$. Here, $\mathcal{P}{z}$ denotes the free-space angular-spectrum propagation over a distance $z$, the distance between the distal facet of the MCF and the target.
As an initial guess of the illumination function, we consider the free-space propagation of the fundamental fiber mode (LP01) of the $i$-th illuminated core from the distal facet of the MCF the target plane. At the target plane, this probe can be approximated by a 2D Gaussian amplitude and a parabolic phase, arises from free-space propagation.

The unknown target object is modeled by a 2D amplitude and phase complex-valued reflectivity profile denoted by $O(\textbf{r})$. The reflected field at the object plane in the $i$-th illumination is given by the product of the illumination (probe) beam and the object reflectivity: $\psi_{i}(\textbf{r}) = O(\textbf{r})\cdot P_{i}(\textbf{r})$. Further simplification can be made by the approximation of a shift-invariant probe function, which perfectly holds for identical MCF cores illuminated at the fundamental modes:  $\psi_{i}(\textbf{r}) = O(\textbf{r})\cdot P_{i}(\textbf{r}) \approx O(\textbf{r}) \cdot P(\textbf{r}-R_{i})$. Where $R_{i}$ is the position shift for the $i$-th illumination core center to its de-scanned position.
The propagation of the reflected field at the target plane, $\psi_{i}(\textbf{r})$, back to the distal facet of the MCF is modeled by the same free-space propagation, giving the complex amplitude of the reflected field at the MCF distal facet:  $\widetilde{\psi}_i (\textbf{r}) = \mathcal{P}_{z}{[\psi_{i}(\textbf{r})]}$. 
This back-reflected field is coupled into the MCF cores, each with a limited collection NA ($\sim NA=0.35$ in our experiments). Due to the spacing between fiber cores ($\sim 4.1 \mu m $ in our experiments) being relatively large compared to the Nyquist criteria ($\sim\lambda/4NA$), the field that is measured at the proximal side by the camera is sparsely sampled.
Thus, the intensity measured at the proximal facet by the camera is modeled by: (i) filtering the field at the distal facet, $\widetilde{\psi}_i (\textbf{r})$ by a Gaussian filter having a bandwidth of $NA \cdot 2\pi/\lambda $ in the angular spectrum domain, (ii) taking the intensity of this 'proximal' field: $I^{prox}_i(\textbf{r}) = |\widetilde{\psi}^{prox}_{i} (\textbf{r})|^2$, (iii) summing the power of this intensity distribution at each core, and assigning this value as the measured intensity in the core position: $I^{cam}_i(\textbf{r}) \approx I^{prox}_i(\textbf{r}) * f_{core}(\textbf{r})$, where $f_{core}(\textbf{r})$ is the core shape, approximated by the average circular core profile.

To reconstruct the complex reflectivity of the object $O(\textbf{r})$ from the intensity patterns measured by the camera, we follow the well-established reconstruction scheme of the ptychographic iterative engine (PIE) \cite{maiden2017further} family, where the propagation between the object plane and the detection plane is given by angular spectrum propagation for a near-field/Fresnel ptychographic reconstruction \cite{stockmar2013near, claus2019diffraction, zhang2019near, xu2020super}. 
At each iteration, the algorithm uses the camera measurements one by one to update the current object and probe estimates such that the probed object diffraction intensity patterns minimize the error with the measured data in a manner similar to hybrid input-output (HIO) conventional phase retrieval \cite{fienup1982phase}. Ptychographic reconstruction offers a more robust reconstruction performance over conventional "keyhole" phase retrieval due to its utilization of joint information from overlapping object parts.
The best reconstruction results were obtained by using the momentum ptychography iterative engine (mPIE) version \cite{maiden2017further}, and the Semi-implicit relaxed Douglas-Rachford algorithm (sDR) whose source codes are available \cite{pham2019semi}, which we adapt to near-field/Fresnel ptychographic scheme using angular spectrum propagation instead of the far-field propagation.

\section{Results}

\subsection{Experimental results}
\begin{figure*}[htb!]
	\centering
	\includegraphics [width=\textwidth]
	{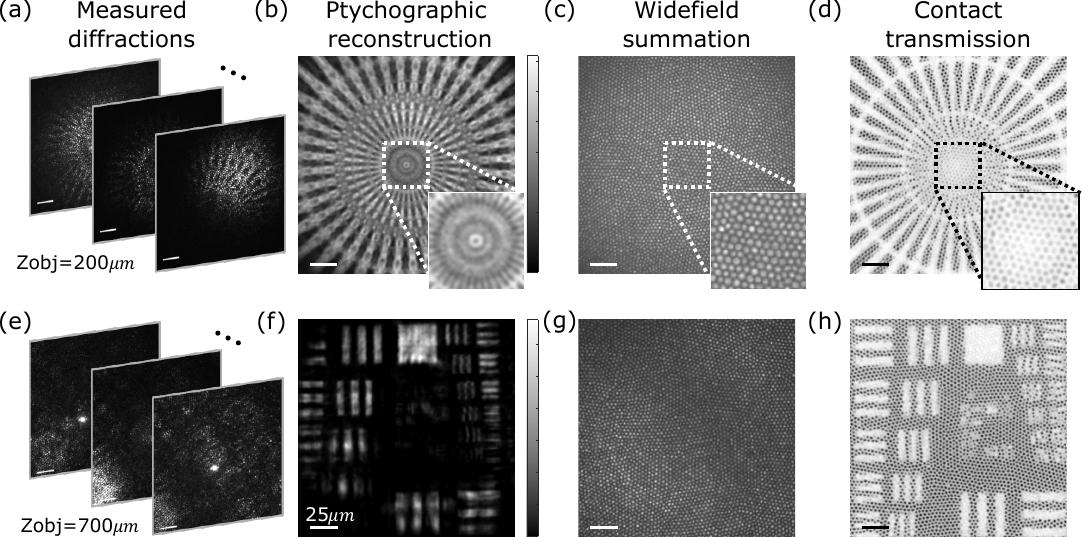}
	\caption{
    \textbf{Experimental imaging of high-resolution targets at different distances from the fiber tip.} (a) a few camera images of the reflected diffraction intensity of a target located 200 $\mu m$ from the fiber tip illuminated from a single core, looking at the proximal fiber facet. (b) From the $N=3000$ intensity patterns, an experimental reconstruction of a Siemens star resolution target is obtained by Fresnel ptychography processing of the raw frames. (c) An image is formed by summing the detected diffraction images in (a), estimating an image given by a wide field illumination. (d) Direct contact measurement of the target, back-illuminated by an incoherent illumination. Insets in (b-d) show magnified images of the target center, demonstrating resolution improvement over widefield summation and contact images. (e-h) Same as (a-d) for the USAF resolution target 700 $\mu m$ from fiber tip and $N=2000$ intensity patterns. Scale bars: 25$\mu m$. %%Maybe necessary to show the groundtruth images?
    	}
	\label{fig2}
\end{figure*}

\begin{figure*}[htb!]
	\centering
	\includegraphics [width=\linewidth,]
	{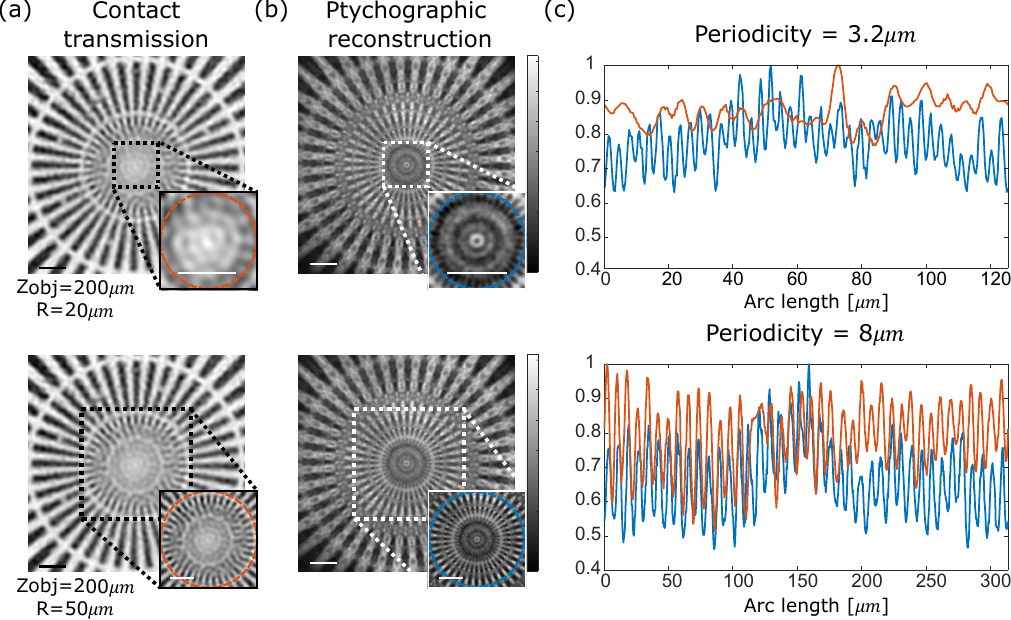}
        \caption{\textbf{Experimental resolution demonstration using a Siemens star target with 40 spoke and space pairs from 150 $nm$ width.} (a) Direct contact measurement with incoherent back illumination featuring a high-resolution spoke zoom-in. (b) Our lensless ptychographic endoscopy reconstruction of the target, placed 200$\mu m$ away from the fiber tip, with a matching zoom-in. (c) Amplitude plots along the circumference of the red circle in (a) and the blue circle in (b) at 3.2$\mu m$ and 8$\mu m$ periodicity, respectively, demonstrate our method resolving pairs that are indistinguishable in contact measurements at 3.2 $\mu m$ periodicity. Scale bars: 25$ \mu m$}
        \label{fig3}
\end{figure*}

Our experimental setup is schematically depicted in Fig.~\ref{fig1}a, with the experimental results summarized in Fig.~\ref{fig2} and Fig.~\ref{fig3}. In all experiments, a diode laser (Finesse Pure, Laser Quantum) with a wavelength of $\lambda = 532 \:nm$ is focused and reflected by a two-axis galvanometer mirror (GM) onto the proximal facet of a one-meter-long MCF  (Fujikura, FIGH-10-350S), having $\sim 8,350$ cores, each with an average core diameter of $\sim$ 2 $\mu m$, and an average core-to-core distance of 4.1 $\mu m$\, and a total diameter of 350 $\mu m$ \cite{choi2022flexible, chen2008experimental}.

As the target objects, we used reflective test targets (Ready Optics Siemens Star, Edmund Positive 1951 USAF Target) placed in a water-filled tank at known distances ranging from $100 \mu m$ to $700 \mu m$ from the distal facet. During each acquisition step, a single core is illuminated while the neighboring cores of the same fiber collect the back-reflected signals. The illumination beam's numerical aperture (NA) is chosen so that at each illuminated core, the illumination mainly excites the fundamental fiber mode (LP01). This ensures that the illumination probe at the object is similar throughout the scan. 
To take into account the possible variation of the probe beam from the ideal fundamental fiber mode, we allow the reconstruction algorithm to correct the initial probe guess during the iterations \cite{thibault2009probe}. 
While core-to-core variations in excitation and core shape are present \cite{chen2008experimental}, allowing the algorithm to update the probe separately at each illumination position did not improve the reconstruction quality.

During the acquisition, the sCMOS camera (pco. edge 4.2) captures an image of the optical intensity at the MCF proximal facet, magnified by a factor of $\sim 37$. As mentioned above, the MCF cores undersample this intensity distribution due to the inherently large core-to-core distance compared to the core diameter. To accommodate this undersampling, we have used as the input to the algorithm for each illumination an intensity distribution that is an interpolation of the raw measured cores intensities.

%%%%%%%%%%%%%%% Figure 2 - Amplitude results %%%%%%%%%%%%%%%%%%%%%%

Sample reconstructions of the different targets are presented in Figures \ref{fig2}-\ref{fig3}. Fig.~\ref{fig2} presents the reconstruction of a Siemens star resolution target and a USAF target at distances of $200 \mu m$ and  $700 \mu m$, respectively. For these reconstructions, a positivity constraint was used in the object plane. We study the effect of different object-plane constraints in Supplementary Section 3.
Several raw measured camera frames are given in Fig.~\ref{fig2}a,e, demonstrating that while the diffracted patterns at close distances show some similarity with the illuminated object parts (Fig.~\ref{fig2}a), the reflected patterns at the longer distance from the sample (Fig.~\ref{fig2}e) are diffracted over a large area, with an average intensity that is significantly lower than the reflection from the illuminated core facet. 
We use $2,000-3,000$ frames of the proximal facet as an input to the Fresnel Ptychographic algorithm to reconstruct the objects (Fig.~\ref{fig2}b, f). The objects can be reconstructed from fewer illuminations, as we study in Supplementary Section 2. 
While the Ptychographic reconstructed images clearly show the fine details of the targets, conventional widefield images taken through the same MCF (by summing the measured patterns) are unable to resolve any object features (Fig.~\ref{fig2}c, g), due to the diffraction from free-space propagation back to the fiber distal facet. We note that the relatively strong reflections from the illuminated core (Fig.~\ref{fig2}e) do not allow confocal-reflection endoscopy. We have digitally filtered the illuminated cores from the raw measured images to mitigate these strong reflections.

Reference contact measurements of the two targets were taken by bringing the tip of the fiber bundle into contact with each target and illuminating the targets from the back using a light-emitting diode (Thorlabs M530L3). These reference images are shown in Fig.~\ref{fig2}d,h, and represent the best images that can be achieved by conventional MCF endoscopy.
However, unlike conventional MCF-based micro-endoscopes where each fiber core acts as an imaging pixel, limiting the spatial resolution to twice the core-to-core pitch (Fig.~\ref{fig3}a), advanced holographic \cite{choi2022flexible, badt2022real} and wavefront-shaping imaging approaches \cite{andresen2013toward} can theoretically offer diffraction-limited resolution and unpixelated field reconstruction (Fig.~\ref{fig3}b). This feature is also provided by our ptychographic endoscopy approach, as can be already observed by comparing the contact reference images (Fig.~\ref{fig2}d,h) and the ptychographic reconstructions (Fig,2b,f). To further study and demonstrate this feature, we provide in Fig.~\ref{fig3} two azimuthal cross-sections through the Siemens star target at radii of $20$ and $50 \mu m$, corresponding to periodicity frequencies of $3.2$ and $8 \mu m$, respectively (Fig.~\ref{fig3}c). Our ptychographic reconstruction improves even over conventional contact measurements. It can distinguish line pairs at a $3.2 \mu m$ and $8 \mu m$ periodicity, while the conventional contact measurement can only separate spokes with an $8 \mu m$ periodicity.

%%%%%%%% Figure S3 - resolution %%%%%%%% 

\subsection{Numerical study of complex (amplitude and phase)  targets}

\begin{figure}[htb!]
	\centering
	\includegraphics [width=\textwidth,]
	{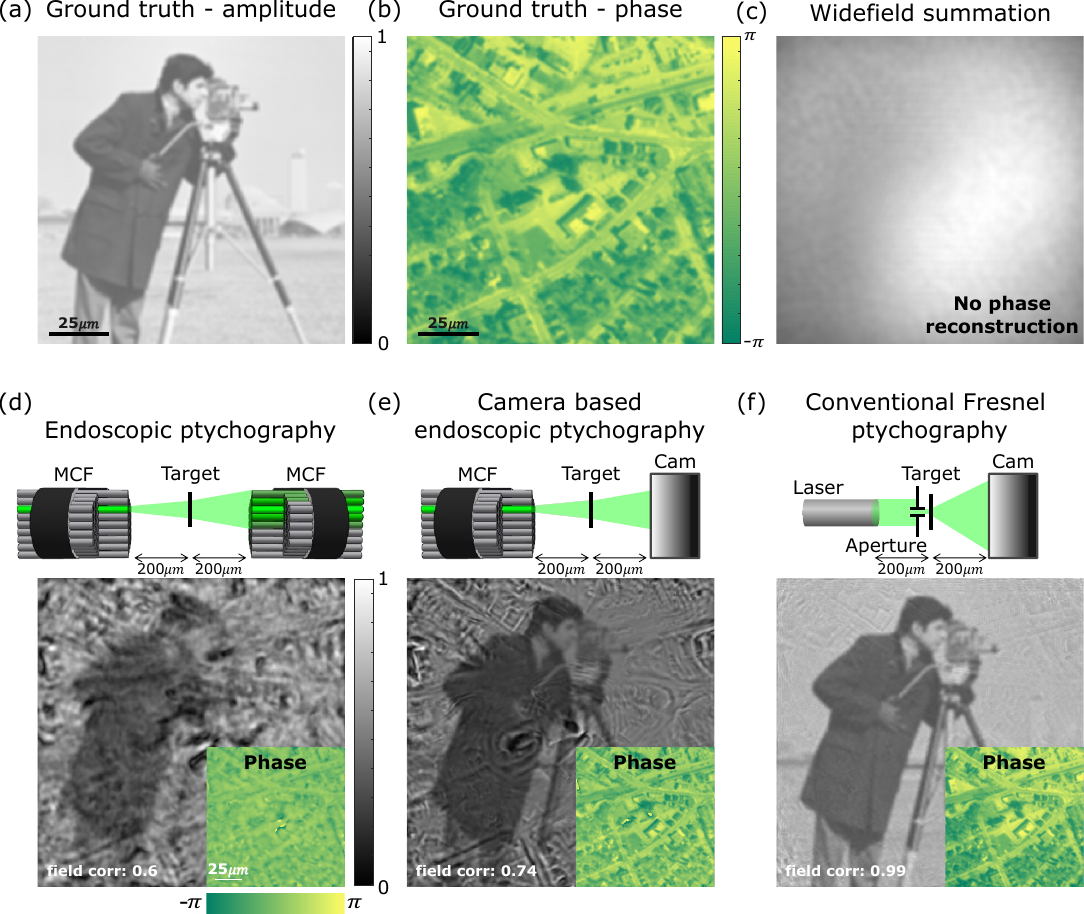}
	\caption{
        \textbf{Numerical investigation of complex (Amplitude and Phase) targets.} (a-b) An object with complex amplitude and phase patterns is placed at 200 $\mu m$ from the fiber tip. (c) Widefield summation of the target cannot recover the amplitude or phase of the target. (d) Reconstruction using our suggested lensless ptychographic endoscopy is given from $N=841$ different core illuminations. The detection scheme we suggest is insufficient for a clear reconstruction of the simulated target but shows a significant improvement over the image given in (c). (e) Suggested modification of our setup, switching the detector of the back-reflected intensity patterns to a camera with a high fill factor and higher numerical aperture. (f) Comparison to a conventional ptychography setup \cite{rodenburg2019ptychography}, where an aperture is placed directly against the specimen at each illumination, and the diffraction pattern is captured with a camera. The field correlation coefficient between the simulated object field and each field reconstruction is presented for all simulation results.}
	\label{fig4}
\end{figure}

To further investigate the potential of our method, which can, in principle, recover the phase profile of unlabeled specimens, we numerically simulate different phase and amplitude targets using the same experimental parameters. The results of this study are presented in Fig.~\ref{fig4}.

The simulated object has an amplitude profile given by the 'cameraman' image (Fig.~\ref{fig4}a), and a phase profile that is an orthophoto of West Concord (Fig.~\ref{fig4}b). The simulated object is placed at a distance of $200\mu m$ from the distal facet of the fiber. $841$ cores of the simulated fiber are illuminated sequentially in a Fermat spiral, known to improve the uniformity of illumination overlap in ptychography \cite{huang2014optimization}. A widefield summation, which sums over the $N=841$ detected intensity patterns on the camera, fails to resolve any phase or amplitude information (Fig.~\ref{fig4}c). However, the ptychographic reconstruction (Fig.~\ref{fig4}d) recovers an estimate of the target's amplitude and phase. However, we find that strong ($0-2\pi$) sharp phase modulation results in mixing phase and amplitude information in the reconstructed image. Such artifacts are not present in ideal Fresnel Ptychography (Fig.~\ref{fig4}f). Thus, their presence can be attributed either to the illumination probe being a soft-edge Gaussian with a parabolic phase rather than a top-hat flat-phase probe or to the undersampling of the detected intensity patterns by the MCF cores.

To study the source for these artifacts, we simulate a theoretical endoscopic ptychographic setup that utilizes the same illumination probe but avoids undersampling of the intensity patterns by sampling the diffracted patterns with an ideal camera with a pixel-to-pixel pitch of $0.4\mu$. While this configuration demands major changes, enlarging the device's footprint and requiring angled illumination and detection, we use it to study artifact origins in our suggested setup.
The result of this simulation (Fig.~\ref{fig4}e) shows significantly improved reconstruction, suggesting that the MCF undersampling is a major source of the ptychographic reconstruction artifacts. 
For completeness, we simulate a conventional Fresnel ptychography measurement and reconstruction of the same target using a top-hat flat-phase probe measured by the same camera (Fig.~\ref{fig4}f). The improved reconstruction fidelity of these measurements suggests that the probe shape also affects the ptychographic reconstruction.
We plot the field correlation between the object and each reconstruction to assess the reconstruction quality quantitatively. For clarity, the illumination and detection scheme changes are schematically depicted in a transmission geometry equivalent to unfolding the reflection setup. 

%%%%%%%%%%%%%%% Figure 4 - Phase & Amplitude simulation %%%%%%%%%%%%%%%%%

\section{Discussion}

We have introduced a simple lensless micro-endoscopic approach for coherent reflection imaging, utilizing readily available multi-core fibers. We adopt the image acquisition approach of standard confocal laser microscopy but replace the confocal detection scheme with a camera-based detection and reconstruct the complex-valued object reflectivity by a ptychographic phase-retrieval. %%Add discussion on what determines the achieved resolution?
For imaging fluorescence objects, we have recently proposed a simple reconstruction scheme that employs the principles of image scanning microscopy (ISM) \cite{Weinberg:23}. ISM improves the resolution of conventional MCF-based endoscopes but can not reach the significant resolution improvement of the ptychographic reconstruction, theoretically offering a diffraction-limited resolution.

Since the diffracted patterns span a large area on the MCF for sufficiently far objects, the ptychographic algorithm does not necessarily require scanning the illumination over all MCF cores. In our experiments, the reconstructions were obtained by illuminating approximately only $10-30\%$ of the fiber cores (see Supplementary Section 2). Interesting potential in improving the acquisition speed may be offered by multiplexed measurements \cite{tian2015computational, sidorenko2016single, barolak2022wavelength}.

Our study indicates that the main limitation of the ptychographic endoscopy is in the reconstruction phase and amplitude objects, given the spatial under-sampling of the commercial MCF used. This limitation can be overcome by using higher fill factor fibers or other detector arrays or by changing the reconstruction framework to account for potential priors in the imaged objects. To this end, deep-learning-based techniques may prove useful, as demonstrated in ptychography \cite{kappeler2017ptychnet} and in imaging through diffuser \cite{li2018deep}.

Interestingly, while we have considered only planar reflecting objects here, ptychographic reconstruction can provide a three-dimensional reconstruction of thick samples \cite{godden2014ptychographic, tian20153d, li2018multi}.

\begin{backmatter}
\bmsection{Funding}
\noindent H2020 European Research Council (101002406), Israel Science Foundation (1361/18), and the Institute for Basic Science (IBS-R023-D1).

\bmsection{Acknowledgments}
\noindent We thank A. Levin and N. Badt for helpful discussions and suggestions.

\bmsection{Disclosures}
\noindent The authors declare no conflicts of interest.

\bmsection{Data availability} 
\noindent Data underlying the results presented in this paper are not publicly available at this time but may be obtained from the authors upon reasonable request.

\end{backmatter}

\bibliography{main}
\newpage
\part*{Supplementary Materials}
\author{} %leave this blank
\setcounter{section}{0}
\section{Numerical Simulation parameters}

This section details the parameters used in the simulations of the fiber-scanning techniques presented in Fig.~4 of the main text. The simulation is designed to approximate the differences in reconstruction across various scenarios: a widefield summation, which sums over all detected intensity patterns; our proposed ptychographic endoscopy, using the same fiber bundle for both illumination and detection; a similar endoscope that utilizes fiber bundle illumination but captures back-reflected intensity patterns with a camera (without pixelization artifacts and with a numerical aperture greater than the fiber's NA); and lastly, a conventional ptychography scenario that illuminates the sample with a flat-phase probe with sharp edges, measuring the back-reflected patterns through a camera.

The simulation is performed using the experimental parameters of the fiber bundle used in our experiments (Fujikura, FIGH-10-350S). The simulated fiber bundle average distance between adjacent cores is $4 \mu m$, with an average core diameter of $2 \mu m$ \cite{choi2022flexible, chen2008experimental}.
We presume each fiber core's illumination to be the fundamental mode, modeled using Marcuse's equation \cite{marcuse1977loss} for step-index fibers - a Gaussian beam with a given waist radius: $w = (D/2)\: (0.65\:+\:1.619\:V_{number}^{-1.5}\:+\ 2.879\: V_{number}^{-6}$) where $w$ is the waist radius, $D$ is the core diameter, and $V_{number}$ is the normalized frequency parameter.

As detailed in the 'Forward model and reconstruction algorithm' section of the main text, the simulation is executed as follows: in the $i$-th illumination step, a single fiber bundle core is illuminated in the fundamental mode, $E_{i}(\textbf{r})$. The illumination probe at the target plane is given by the free-space propagation of the fiber mode: $P_{i}(\textbf{r}) = \mathcal{P}{z}{[E_{i}(\textbf{r})]}$, where $\mathcal{P}{z}$ represents the free-space angular-spectrum propagation over a distance $z$ from the fiber facet to the target.
The target object is modeled by a 2D amplitude and phase complex-valued reflectivity profile denoted by $O(\textbf{r})$. The illuminated field at the target plane in the $i$-th, $\psi_{i}(\textbf{r})$, is calculated as the pointwise product of the illumination probe and the object's complex (amplitude and phase): $\psi_{i}(\textbf{r}) = O(\textbf{r}) \cdot P_{i}(\textbf{r})$. We model the probe function to be a shift-invariant probe function, which perfectly holds for identical MCF cores illuminated at the fundamental modes: $\psi_{i}(\textbf{r}) = O(\textbf{r})\cdot P_{i}(\textbf{r}) = O(\textbf{r}) \cdot P(r-R_{i})$. This illuminated field at the target plane, $\psi_{i}(\textbf{r})$, is then back-propagated via the same free-space propagation to the distal facet of the fiber, denoted by $\widetilde{\psi} (\textbf{r})_{i} = \mathcal{P}{z}{[\psi_{i}(\textbf{r})]}$.
The propagation of the reflected field at the target plane, $\psi_{i}(\textbf{r})$, back to the distal facet of the MCF is modeled by the same free-space propagation, giving the complex amplitude of the reflected field at the MCF distal facet: $\widetilde{\psi}_i (\textbf{r}) = \mathcal{P}_{z}{[\psi_{i}(\textbf{r})]}$. 
This back-reflected field is coupled into the MCF cores, each with a collection NA of $NA=0.35$, and modeled as a Gaussian beam. The intensity measured at the proximal facet by the camera is modeled by: (i) filtering the field at the distal facet, $\widetilde{\psi}_i (\textbf{r})$ by a Gaussian filter having a bandwidth of $NA \cdot 2\pi/\lambda $ in the angular spectrum domain, (ii) taking the intensity of this 'proximal' field: $I^{prox}_i(\textbf{r}) = |\widetilde{\psi}^{prox}_{i} (\textbf{r})|^2$, (iii) summing the power of this intensity distribution at each core, and assigning this value as the measured intensity in the core position: $I^{cam}_i(\textbf{r}) \approx I^{prox}_i(\textbf{r}) * f_{core}(\textbf{r})$, where $f_{core}(\textbf{r})$ is the core shape, approximated by a circular profile. The detected intensity is then interpolated using a 'bicubic' kernel to allow propagation without overlapping replicas artifacts. The interpolated intensity images, the probe profile, and the core positions are then input into a Ptychography reconstruction algorithm.

In the scenarios where the back-reflected intensity patterns are captured with a camera, we omit the final steps in which the back-reflected field is convolved with the mode field profile and fiber sampling. Instead, we use a sharp-edged filter with a numerical aperture (NA) of $0.8$.

\newpage{}

\section{Number of illuminations required for image reconstruction}

\begin{figure*}[h!]
	\centering
	\includegraphics [width=\linewidth,]
	{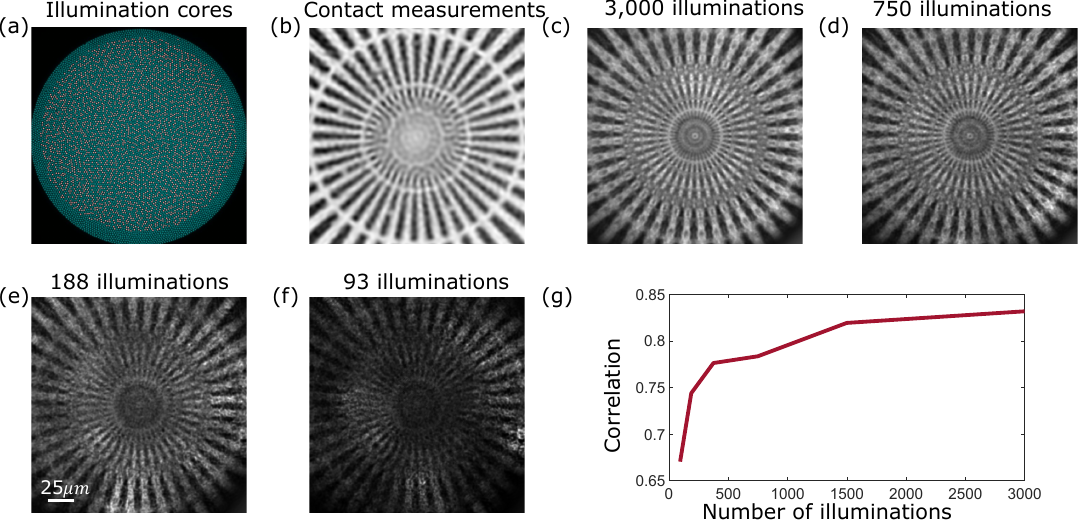}
        \renewcommand{\thefigure}{S1}
	\caption{
            \textbf{Investigation of the reconstruction quality as a function of the number of illuminations used in the experimental data.} (a) Fujikura FIGH-10-350S imaging fiber with approximately 8,350 cores is sequentially illuminated (in red) at 3,000 cores for the ptychographic reconstruction. (b) Contact measurements of a Siemens star resolution target. (c) Conventional widefield image is taken by summing all 3,000 illumination patterns. (d-f) Ptychography reconstruction of the same target in (b), situated at a distance of 100$\mu m$, with 93-3,000 core illuminations out of approximately 8,500 cores of the bundle. (g) Quantification of the reconstruction correlation to the contact measurements. Scale bar: $25\mu m$.
                }
	\label{figS2}
\end{figure*}

While our image acquisition shares similarities with confocal laser endomicroscopy, the reconstruction scheme can work with much fewer cores used for illumination. To investigate the effect of the number of illuminated cores, we provide a reconstruction of the resolution target in Fig.~\ref{figS2} for various numbers of illuminations of the reconstruction target presented in the Main text Fig.~2. We could potentially enhance image acquisition speed by reducing the number of images required for reconstruction.

Supplementary Figure \ref{figS2} shows the reconstruction results with subsets of $93, 188, 750,$ and $3,000$ images. These subsets were selected to ensure maximal overlap between neighboring probes on the object plane, and to guarantee that the entire field of view is illuminated \cite{huang2014optimization}. %all taken for the same field of view (FoV) of the fiber where a full confocal scan would require $\sim 8,350$ images.
We assess the validity of our reconstruction by comparing its correlation with the interpolated contact measurement. This comparison helps to determine the accuracy and effectiveness of the reconstruction process, given different numbers of illuminated cores and images.

%%%%%%%% Figure S2 - reconstruction similarity as a function of n_illum %%%%%%%% 

\newpage
\newpage

\section{Reconstruction constraints}
One strategy for resolving ambiguities in phase-retrieval reconstruction involves the use of additional a priori constraints \cite{rodenburg2019ptychography}. For instance, the assumption of positivity where the object phase is flat, or a phase-only object when the object amplitude is flat, can be used. We demonstrate the impact of using these constraints for flat-phase targets (mirrors) and phase-only targets (with flat amplitude) in both experimental settings and numerical simulations, the same Siemens star target as shown in Figs.~2,3 of the main text, placed at a distance of $100\mu m$ from the distal tip. The numerical phase target is similar to the phase target shown in Fig.~4 of the main text, with a flat amplitude.

These priori constraints allow the ptychographic reconstruction process towards a more precise solution, enhancing the quality of target image reconstruction. This is particularly beneficial when the target possesses known characteristics, such as flat-phase or phase-only properties. The outcomes from both the experimental setups and numerical simulations affirm the efficacy of employing these constraints in enhancing the accuracy of the reconstruction process.

%%%%%%%% Figure S3 - positivity %%%%%%%% 
\begin{figure*}[h!]
	\centering
	\includegraphics [width=\linewidth,]
	{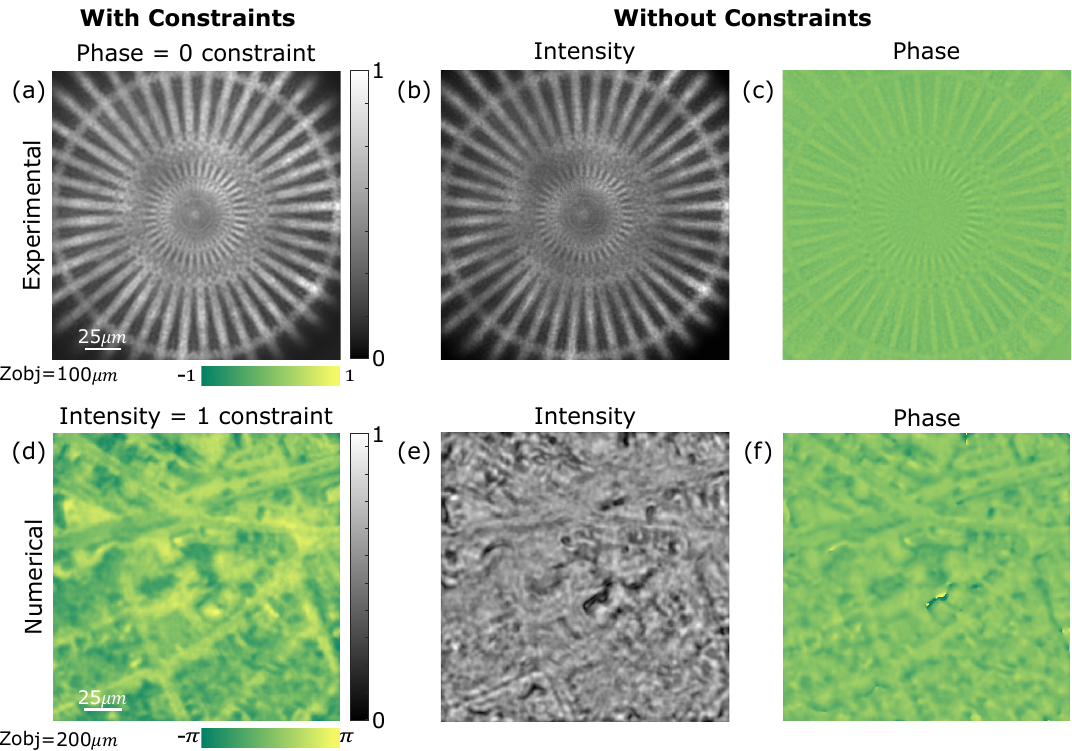}
 \renewcommand{\thefigure}{S2}
	\caption{\textbf{Investigation of reconstruction constraints on the reconstruction quality.} Two test targets are used to compare our ptychographic endoscopy reconstruction with and without the addition of prior knowledge in the form of object-plane constraints. (a) Experimental results of a Siemens test target positioned 100 $\mu m$ from the MCF facet, reconstructed with the positivity constraint. (b-c) Reconstruction uses the same ptychographic algorithm and dataset without the positivity constraint. (d) Numerically simulated phase target with flat amplitude, positioned 200 $\mu m$ from the MCF facet, reconstructed with the constraint that the amplitude is flat. (e-f) Reconstruction uses the same ptychographic algorithm and dataset without the flat amplitude constraint. Scale bar: $25\mu m$.
	}
	\label{figS3}
\end{figure*}

\end{document}